\documentclass[11pt,twoside]{article}
\usepackage{asp2014}

\aspSuppressVolSlug
\resetcounters

\begin{document}

\title{Post-Asymptotic Giant Branch Evolution of  Low- and Intermediate-Mass Stars. Preliminary Results.}
\author{Marcelo M. Miller Bertolami$^{1,2,\dagger}$ 
\affil{$^1$Max-Planck-Institut f\"ur Astrophysik, Karl-Schwarzschild-Str. 1, 85748, Garching, Germany; \email{marcelo@MPA-Garching.MPG.DE}}
\affil{$^2$Instituto de Astrof\'isica de La Plata, UNLP-CONICET, Paseo del Bosque s/n, 1900 La Plata, Argentina; \email{mmiller@fcaglp.unlp.edu.ar}}}
\affil{$^\dagger$Postdoctoral fellow of the Alexander von Humboldt Foundation}
\paperauthor{M. M. Miller Bertolami}{marcelo@MPA-Garching.MPG.DE}{}{Max Planck Gesellschaft}{Max-Planck-Institut f\"ur Astrophysik}{Garching}{Bayern}{85748}{Germany}

\begin{abstract}
Preliminary results from an ongoing project to compute a grid of post-AGB
models is presented. Our preliminary results show that stellar evolution
computations that include an updated treatment of the microphysics predict
post-AGB timescales that are several times shorter that predicted by older
models. Also the mass-luminosity relation of post-AGB models deviates from
that of older grids. In addition, our results suggest only a slight
metallicity dependence of the post-AGB timescales. We expect these results to
have significant consequences for models of the formation of planetary
nebulae and their luminosity function.
\end{abstract}

\section{Introduction}
Planetary Nebulae (PNe) are among the most beautiful astronomical
objects. They are the result of the evolution of low- and intermediate-mass
stars ($M_{\rm ZAMS}\sim 0.8-8\ M_\odot$). In the most simple scenario, PNe
are formed when the progenitor stars lose their external envelopes at the end
of the Asymptotic Giant Branch (AGB) and cross the HR-diagram on their
way to the white dwarf cooling sequence. While crossing the HR-diagram the
central stars of the PNe (CSPNe) become sufficiently hot to ionize the
previously ejected material \citep{1957IAUS....3...83S,1966PASP...78..232A,1970AcA....20...47P}.

Besides being interesting and fascinating objects in themselves their
properties are also useful for other fields of astrophysics
\citep{2014arXiv1403.2246K}. PNe and CSPNe offer unique insight into the
nucleosynthesis during previous evolutionary phases like the
AGB. Extragalactic PNe can be used to understand metallicity gradients and
their temporal evolution in galaxies. Also, the PNe luminosity function (PNLF)
has proven to be a good distance indicator as far as $\sim 20$Mpc, but we
still do not understand why \citep{2012Ap&SS.341..151C}.  The formation and
detectability of PNe depends strongly on the relationship between two
different timescales. The evolutionary timescale of the CSPNe, which provides
the ionizing photons, and the dynamical timescales of the circumstellar
material ejected at the end of the AGB. If the CSPN evolves too fast the PN
will be ionized for a short time, and thus will have a low detection
probability, or might even not be ionized at all. On the other hand, if the
star evolves too slowly, the ionization of the nebula will take place when the
ejected material has already dispersed too much to be detectable.  In this
work we address the first of these timescales. Namely, we present preliminary
results from full stellar evolution computations of the post-AGB and CSPNe
phases that include an updated treatment of macro and microphysics. This stage
is one of the least understood phases of low- and intermediate mass stars and
there are some indications that current models are not accurate enough,
e.g. a) The two available grids of post-AGB models \citep{1994ApJS...92..125V,
  1995A&A...299..755B} do not agree on the predicted timescales
\citep{2008ApJ...681.1296Z}, b) The CSPNe mass-luminosity relation seems to be
at variance with the constraints coming from hydrodynamically consistent model
atmospheres \citep{2004A&A...419.1111P}, c) consistency between the masses of
white dwarfs and those of CSPNe determined by asteroseismology requires faster
evolutionary speeds \citep{2014A&A...566A..48G}. In addition, until now we do
not understand why the cut-off of the PNe luminosity function is constant in
most galaxies. Last, but not least, \cite{2002A&A...387..507M} showed that
C-rich molecular opacities are key to predict the right effective temperatures
once the AGB models become carbon rich ($C/O>1$, by number fraction). These
inconsistencies and the fact that available post-AGB models are based on very
old radiative opacities and microphysics calls for a restudy of the problem.

\section{Physical details of the stellar evolution models}
The computations presented in this work have been performed with {\tt LPCODE}
stellar evolution code. A detailed description of the code can be found in
\cite{2013A&A...557A..19A}, and references therein, here focus on the physical
ingredients, and updates, which are particularly relevant for the present
work. In the pre-WD regime {\tt LPCODE} uses the OPAL EOS\_2005 equation of
state for H- and He rich mixtures and a simplified EOS for other
compositions. Updated high- and low- temperature radiative opacities are
included according to \cite{1996ApJ...464..943I} and
\cite{2005ApJ...623..585F}. This includes pretabulated C-rich molecular
opacities \citep{2009A&A...508.1343W}. The $^{14}$N(p,$\gamma$)$^{15}$O
reaction rate, that sets the overall efficiency of the CNO-cycle, was taken
from \cite{2005EPJA...25..455I}. Convective mixing is treated within mixing
length theory (MLT) and a diffusive convective picture, including an
exponentially decaying velocity field outside formal convective boundaries
(with a free parameter $f$, see \citealt{1997A&A...324L..81H} for
details). From the calibration of the solar model with diffusion we obtain
$\alpha_{\rm MLT}=1.825$. The value of $f$ in convective cores is set to
$f=0.0174$ from the calibration of the width of the upper main sequence. This
equivalent to an overshooting extension of 0.2 times the pressure scale
height. The values of $f$ in the pulse driven convection zone (PDCZ) during
the thermal pulses in the AGB is set to $f^{\rm PDCZ}=0.005$ which allows to
reproduce the range of He, C and O ratios observed in PG1159 type stars
\citep{2006PASP..118..183W}. The $f$-value at the bottom of the convective
envelope is taken to be $f^{\rm CE}=0.1$ \citep{2005ARA&A..43..435H}. Winds
are a decisive aspect of AGB evolution since they determine when the TP-AGB
phase ends. To include the impact of the transition from an O-rich AGB to a
C-rich AGB star we implemented different wind prescriptions for the O- and
C-rich dust driven winds. For the sake of consistency we choose dust driven
wind prescriptions derived by the same authors and methods, ($\log
\dot{M}=4.08\, \log P-16.54$ \citealt{1998MNRAS.293...18G} and $\log
\dot{M}=-9+ 0.0032\, P$ \citealt{2009A&A...506.1277G}; where $P$ is the
pulsation period). For the pre-dusty winds we included the
\cite{2005ApJ...630L..73S} which seems to reproduce some RGB and AGB
observables better than the standard Reimers prescription
\citep{2010ApJ...724.1030G}. Finally, for the hot radiative driven winds we
adopted a mass loss prescription, $\dot{M}=9.8\times 10^{-15}\times
(L/L_\odot)^{1.674}$, which is based on the results of
\cite{2004A&A...419.1111P} and similar to the one adopted by
\cite{1995A&A...299..755B}. Between the the hot and cold wind regime,
$3.8\lesssim \log T_{\rm eff}\lesssim 4.1$, there are no available
prescriptions and we had to rely on interpolations.

\section{Preliminary results and discussion}
\begin{figure}[ht]
\includegraphics[clip, angle=0, width=13.cm]{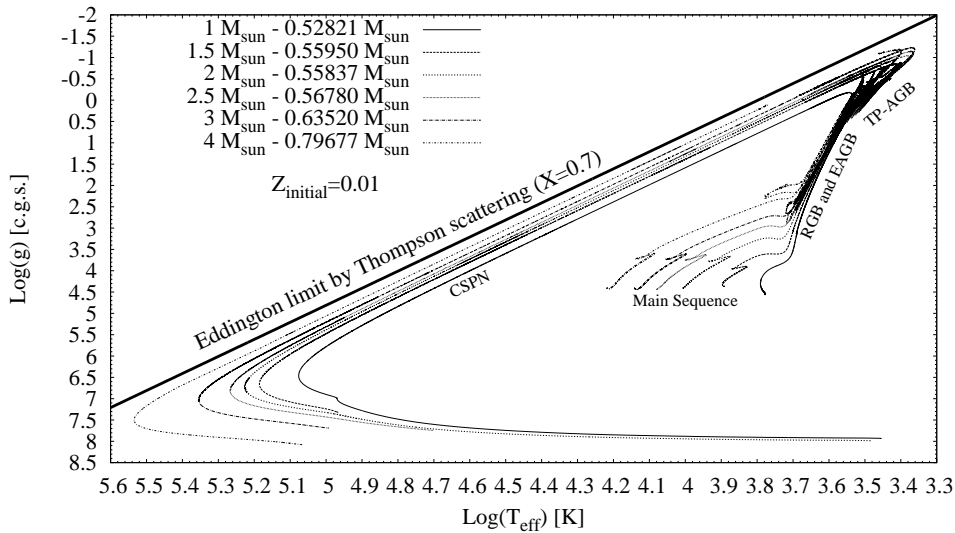} 
\caption{$\log T_{\rm eff}-\log g$-diagram of the computed sequences for Z=0.01.}
\label{Fig:Kiel}
\end{figure}
\begin{figure}[ht]
\includegraphics[clip, angle=0, width=13.cm]{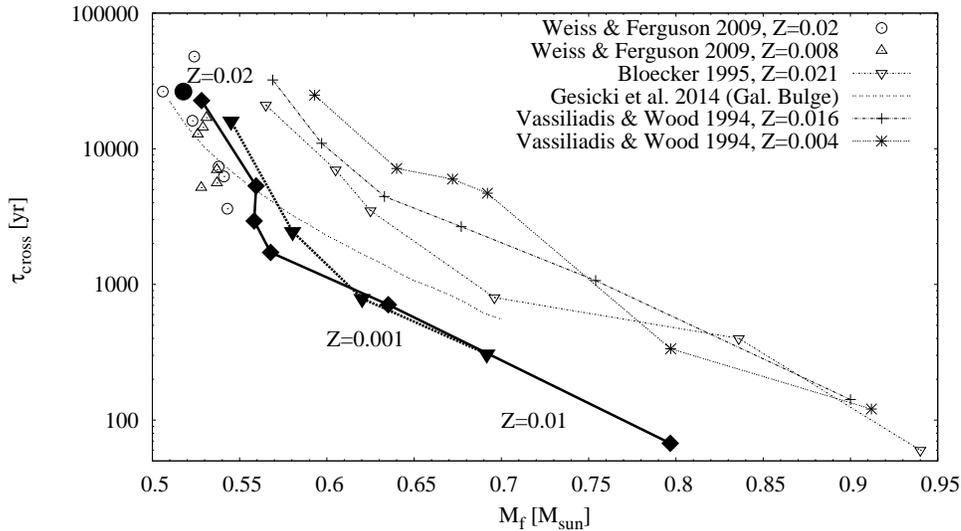}  
\caption{Crossing times ($\tau_{\rm cross}$) from different sources as
  compared with our results for H-burning post-AGB sequences. Filled symbols
  indicate the predicted crossing times for Z=0.02, 0.01, 0.001; circles,
  rhombi and triangles respectively.}
\label{Fig:Crossing}
\end{figure}
\begin{table}
\caption {Properties of the AGB and post-AGB (H-burning) stellar evolution models: metallicity ($Z$), initial mass ($M_i$), final mass ($M_f$),
  number of thermal pulses on the AGB ($N_{\rm TP}$), age of the model at the
  first thermal pulse ($\tau_{\rm 1TP}$), length of the O-rich TP-AGB
  ($\tau_{O}$), length of the C-rich TP-AGB ($\tau_{C}$), the luminosity of
  the post-AGB remnant in the plateau phase ($L^{\rm post-AGB}_{\log T_{\rm
      eff}=4}$, taken at $\log T_{\rm eff}=4$), and the crossing time
  ($\tau_{\rm cross}$) of the post-AGB remnant form $\log T_{\rm eff}=4$ to
  the point of maximum effective temperature (``knee'', see
  Fig. \ref{Fig:Kiel}).}
\label{tab:grid} 
\begin{center}
   \begin{tabular}{ c c | c  c  c c c c c} 
 $Z$ & $M_i$ & $M_f$ & $N_{\rm TP}$ & $\tau_{\rm 1TP}$ & $\tau_{O}$ &$\tau_{C}$ & $L^{\rm post-AGB}_{\log T_{\rm eff}=4}$ & $\tau_{\rm cross}$ \\ 
     & [M$_\odot$]&[M$_\odot$] & & [Myr] &  [Kyr] & [Kyr] & [$L_\odot$] &[yr] \\ \hline
 0.02  & 1.0 & 0.51781 & 3  & 11923 & 626$\lesssim$ &   0  & 2966 & 26407  \\
 0.01  & 1.0 & 0.52821 & 4  & 10510 & 728$\lesssim$ &   0  & 3396 & 22660  \\
 0.01  & 1.5 & 0.55950 & 7  & 2584  & 814 & 260$\lesssim$  & 5674 & 5319 \\
 0.01  & 2.0 & 0.55837 & 12 & 1206  & 1513& 654$\lesssim$  & 6624 & 2936 \\
 0.01  & 2.5 & 0.56780 & 11 & 720   & 611 & 1105$\lesssim$ & 7700 & 1722 \\
 0.01  & 3.0 & 0.63520 & 8  & 431   & 177 & 345$\lesssim$  & 10524 & 708 \\
 0.01  & 4.0 & 0.79677 & 11 & 195   & 28  & 109$\lesssim$  & 18270 & $\sim 67$ \\
 0.001 & 1.0 & 0.54510 & 4  & 6594  & 625 & 249$\lesssim$  & 4477 & 15983 \\ 
 0.001 & 1.5 & 0.58039 & 7  & 1754  & 0   & 1017$\lesssim$ & 8100 & 2454  \\ 
 0.001 & 2.0 & 0.62018 & 10 & 797   & 82  & 774$\lesssim$  & 11434 & 787   \\ 
 0.001 & 2.5 & 0.69163 & 12 & 492   & 0   & 570$\lesssim$  & 14920 & 307   \\ 
   \end{tabular}
\end{center}
\end{table}

It is well known that the computation of the very end of the TP-AGB is riddled
with convergence problems \citep{2009A&A...508.1343W,2012A&A...542A...1L}.
This implies that a lot of human time (baby-sitting) is required to compute
the transition from the TP-AGB to the CSPNe phase. Even when codes converge,
convergence happens at the expense of a prohibitively small timesteps (even
down to $\Delta t\sim 1$ hour). Despite all numerical improvements our
computations are not the exception. Consequently, the number of complete
simulations presented in this work is rather small. In Fig. \ref{Fig:Kiel} we
show the $\log T_{\rm eff}-\log g$-diagram (``Kiel diagram'') for our $Z=0.01$
sequences. Table \ref{tab:grid} shows the most relevant quantities of our
TP-AGB and post-AGB stellar models. A comparison of the $M_i$-$M_f$
relationship of our models (Table \ref{tab:grid}) with semiempirical
determinations from stellar clusters \citep{2009ApJ...692.1013S}, common
proper motion pairs \citep{2008A&A...477..213C} or the Galactic bulge
\citep{2014A&A...566A..48G} suggest that our values of $M_f$ may me somewhat
low. This is very likely the consequence of too strong third dredge up during
the TP-AGB \citep{2009ApJ...692.1013S}. In fact, this causes the formation of
C-rich stars already at the first thermal pulse in our low metallicity
sequences (Z=0.001 and $M_i=1.5$ and 2.5 M$_\odot$). This might lead to
disagreements with the carbon star luminosity function and thus to a need to
re-calibrate the overshooting parameters during the TP-AGB. However, it should
be notice that the value of $f^{\rm PDCZ}\sim 0.005$ is required to reproduce
the abundances of PG1159 type post-AGB stars, leaving only $f^{\rm CE}$ as a
possible free parameter.

A comparison of our timescales with those of previous models and semiempirical
determinations (see Fig. \ref{Fig:Crossing}) allow us for some very
interesting preliminary conclusions. On one hand, in the range of remnant
masses where our models overlap with those of \cite{2009A&A...508.1343W} the
agreement is quite good. This is particularly interesting in the light of the
different numerical codes and wind prescriptions adopted. This suggests that
the details of mass loss might not play a crucial role in the determination of
post-AGB timescales. Given the uncertainties behind this ingredient of stellar
evolution computations, this might be a good news. On the other hand, our
models predict much shorter (up to a factor of $\sim 5$!) post-AGB timescales
than those of \cite{1994ApJS...92..125V} and \cite{1995A&A...299..755B}. In
the light of the agreement between our timescales and those of
\cite{2009A&A...508.1343W} this suggests that stellar evolution computations
based on old microphysics might have significantly overestimated the length
of the CSPNe phase. Interestingly enough our much shorter timescales are in
agreement with the results from studies of PNe
\citep{2007A&A...467L..29G,2014A&A...566A..48G} that suggests that CSPNe
should evolve several times faster than predicted by old stellar evolution
models. Last, but not least, our models do not predict a strong dependence of
the post-AGB timescales with metallicity. In addition, the post-AGB
mass-luminosity relation of modern models is different from that of old
grids. All these results, if confirmed, will have an impact in the predictions
of models for the formations of PNe in different stellar populations. In
particular, the impact of modern post-AGB computations in the formation of the
PNLF needs to be assessed.


\end{document}